\documentclass[a4paper,oneside,12pt]{elsarticle}
\usepackage{amsthm}
\usepackage{amsmath}
\usepackage{graphicx}%
\usepackage{amsfonts}%
\usepackage{amssymb}
 \usepackage{verbatim}
 \usepackage{float}
\usepackage{rotating}
\usepackage{amsmath,amsthm,amssymb}

\begin{document}

\author{Sovan Mitra}
\title{\huge{\textbf{An Introduction to Hedge Funds}}} \maketitle
\section*{\large{Abstract}}
This report was originally written as an industry white paper on
Hedge Funds. This paper gives an overview to Hedge Funds, with a
focus on risk management issues. We define and explain the general
characteristics of Hedge Funds, their main investment strategies
and the risk models employed. We address the problems in Hedge
Fund modelling, survey current Hedge Funds available on the market
and those that have been withdrawn. Finally, we summarise the
supporting and opposing arguments for Hedge Fund usage.

A unique value of this paper, compared to other Hedge Fund
literature freely available on the internet, is that this review
is fully sourced from \textit{academic} references (such as peer
reviewed journals) and is thus a bona fide study.

This paper will be of interest to: Hedge Fund and Mutual Fund
Managers, Quantitative Analysts, ``Front" and ``Middle" office
banking functions e.g. Treasury Management, Regulators concerned
with Hedge Fund Financial Risk Management, Private and
Institutional Investors, Academic Researchers in the area of
Financial Risk Management and the general Finance community.
\\\\
\textbf{Key words}: Hedge Funds, risk management, risk
measurement, regulation.

\line(1,0){400}

\section{Introduction and Outline}
According to the European Central Bank \cite{GARBECB}, the Hedge
Fund industry is growing rapidly with a total of US \$1 trillion
worth of assets under their control globally. A Hedge Fund's size
is typically less than US \$100 million, with nearly half under US
\$25 million \cite{GARBECB}. They represent a small percentage of
the asset management industry (see \cite{GARBECB}) yet they exert
a disproportionately massive influence on the financial and
economic sector in relation to their size (see Fung
\cite{FUNGImpact}). This is due to Hedge Funds generally using
dynamic and leveraged trading strategies, which is in contrast to
Mutual Funds that typically engage in buy-and-hold strategies.
Thus it is apparent Hedge Funds have a significant influence in
financial markets, yet knowledge of them is relatively little.
\\\\
In this paper we introduce Hedge Funds, attempting to firstly
propose a definition for Hedge Funds as no common consensus has
yet been agreed within the Finance community. We then explain the
common investment strategies applied by Hedge Funds e.g. event
driven, long-only investment. In the next section, we survey the
main risk models applied to analysing Hedge Funds whilst also
discussing the difficulties in actually measuring Hedge Fund
risks. Finally we finish by surveying current Hedge Funds
available on the market and famous Hedge Funds that have been
withdrawn.
\\\\
It is important to note that knowledge and performance of the
Hedge Fund industry is guarded with substantial secrecy.
Consequently, the quality of information used in any Hedge Fund
study, can never be as good as those for other investment products
e.g. Mutual Funds (see Fung \cite{FUNGImpact},Fung
\cite{FUNGPrimer}, Do et al. \cite{DOHedge}).

\section{Introduction to Hedge Funds}
Within the investment industry, many fund types exist: Hedge
Funds, investment trusts, unit trusts etc... yet the term Hedge
Fund has no explicit definition. In fact the European Central Bank
states in its report on Hedge Funds \cite{GARBECB} that no common
Hedge Fund definition exists. Defining a Hedge Fund is in fact
more problematic than it appears. To appreciate the difficulty in
defining a Hedge Fund, it is instructive to know its brief
history.

\subsection{Brief History of the Hedge Funds Industry}
 According to Fung
\cite{FUNGPrimer}, the first ever Hedge Fund was formed by Albert
Wislow Jones in 1949, so called as the main investment strategy
was to take hedged equity investments. By hedging (the act of
removing risk in some investment by taking an investment in
another (typically related) investment) Winslow was able to
eliminate some market risks.
\\\\
Hedge Funds then became first well-known after an article in
\textsl{Fortune}(1966) mentioning Jones's fund significantly
outperforming other Mutual Funds \cite{FUNGPrimer}. Although this
article initiated wide interest in Hedge Funds, their popularity
diminished as it fell victim to the bear markets of 1969-70 and
1973-4. A decade later (1986), interest was revived by Robertson's
infamous Tiger Fund \cite{FUNGPrimer}, which achieved compound
annual returns of 43\% for 6 years after all expenses. Fung in
\cite{FUNGPrimer} corroborates the impact that the publicity of
Robertson's Fund had on the Hedge Fund industry by showing the
rapid expansion of Hedge Funds and CTA funds (commodity trading
advisor funds (similar to Hedge Funds)) from 1985-97.
\\\\
With numerous Hedge Fund investors and the fact that Hedge Funds
were virtually unregulated compared to other funds, a multitude of
new Hedge Fund trading strategies evolved, including the use of
derivatives e.g. options. Now all these funds came to be known as
Hedge Funds yet many of them were using investment strategies
beyond simply ``hedging" that A.Winslow first employed (see
\cite{GARBECB} for more details). To complicate matters further,
as Hedge Fund strategies developed so also did funds other than
Hedge Funds begin employing Winslow's equity hedging strategy,
thus hedging was no longer unique to Hedge Funds. Today, the word
``hedge" in Hedge Funds has become a misnomer, more of a
historical hangover from Alfred Winslow rather than a description.

\subsection{A Definition of Hedge Funds}
 As the European Central bank states \cite{GARBECB}:
 \\\\
\textit{``there is no common definition of what constitutes a
Hedge Fund,it can be described as an unregulated or loosely
regulated fund which can freely use various active investment
strategies to achieve positive absolute returns"}.
\\\\
As the European Central Bank implies, a Hedge Fund is difficult to
define partly because of a lack of clarity of agreement on its
term and also due to its diverse trading spectrum. They are
typically characterised by high leveraging, derivatives trading
and short selling compared to Mutual Funds. One way of defining a
Hedge Fund is by comparing the similarities and differences with
Mutual Funds. In a sense Hedge Funds are similar to any other
portfolio investment in 3 respects:
\begin{itemize}
\item they are funded by capital from investors, rather than bank
loans or other sources of capital;
\item they invest in publicly traded securities e.g. equities and
bonds;
\item the capital is ``managed" or invested by expert fund
managers.
\end{itemize}

The key differences between Hedge Funds and Mutual Funds lies in
the degree of regulation, the level and variety of risky
investment strategies. Whereas Mutual Funds are required to adhere
to strict financial regulations, including the types and levels of
risks, Hedge Funds are free to pursue virtually any investment
strategy with any level of risk.
\\\\
Secondly, Hedge Fund investors are typically high net worth
individuals or institutional investors e.g. pension funds
\cite{GARBECB}, partly because Hedge Funds typically require high
minimum investment amounts. A graph taken from the European
Central Bank \cite{GARBECB} shows the composition of Hedge Fund
investors from 1992-2004.
\\\\
\includegraphics[height=70mm,width=90mm]{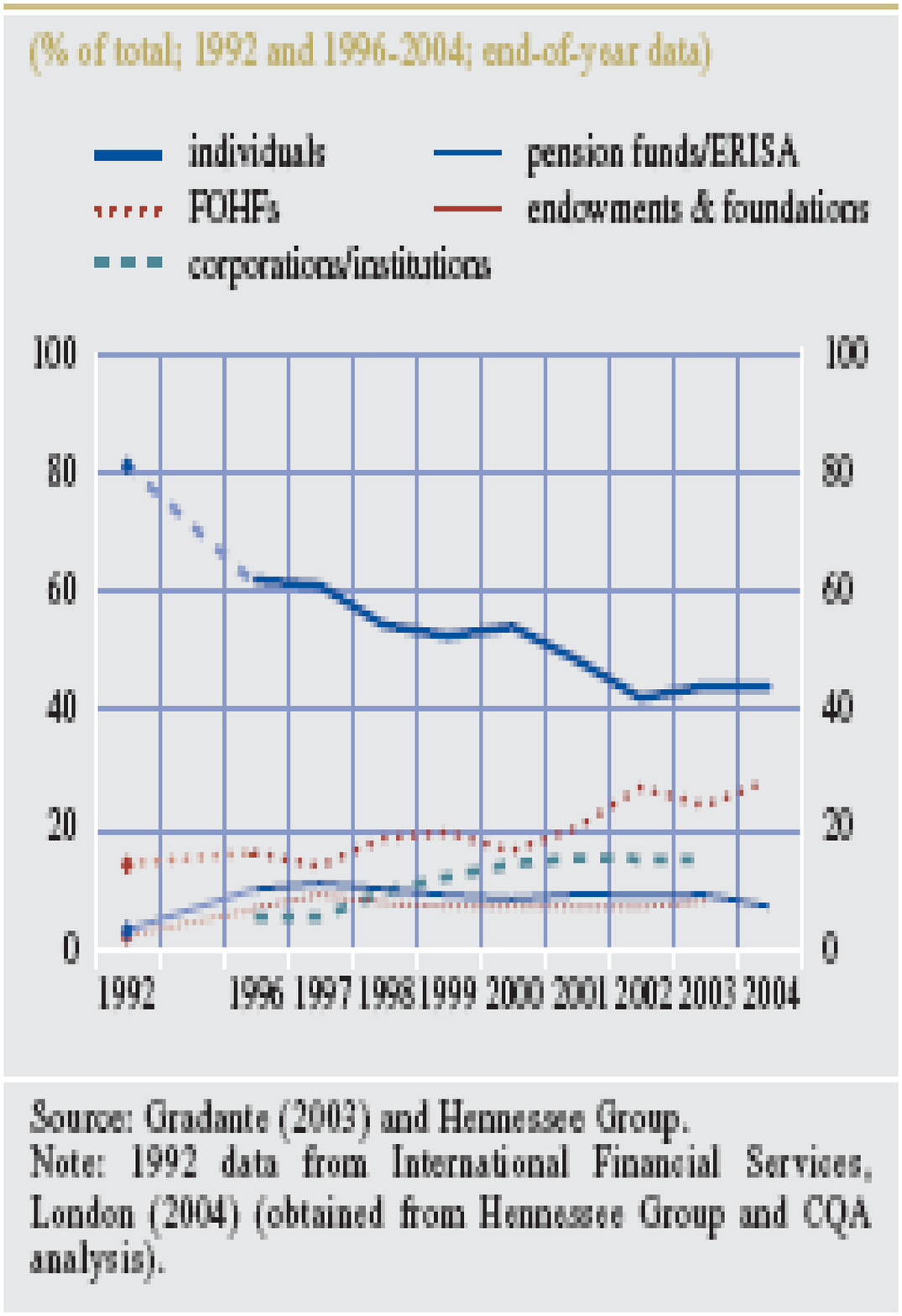}
\\\\
Mutual funds on the other hand, are typically targetted at the
general public and will accept any investor who can meet the
minimum investment amount. Hedge Funds in fact are banned from
advertising and in some cases the investors are required to be
``accredited".
\\\\
A third key difference is the fund portfolio composition. As Fung
\cite{FUNGPrimer} states, the majority of Mutual Funds are
composed of equities and bonds. Hedge fund portfolio compositions
are far more varied, with possibly a significant weighting in
non-equity/bond assets e.g. derivatives.
\\\\
A fourth key difference is that the historical return
characteristics and distribution of Hedge Funds tend to differ
significantly from Mutual Funds. For example, Capocci et al.
\cite{CAPOAnal} and Getmansky \cite{GETMEco} demonstrate that
Hedge Funds empirically display serial correlation in returns.
According to Brown \cite{BROWVol}, Hedge Funds do not perform
significantly better than most investment funds; Hedge Funds
between 1989-95 earned 300 basis points below the S\&P 500.
However, other studies conclude that Hedge Funds produce excess
market returns (see \cite{CAPOAnal},\cite{DOHedge}). A graph below
from \cite{GARBECB} gives the performance of Hedge Funds compared
to key indexes. The CSFB/Tremont index is a Hedge Fund index, the
``equivalent" of the FTSE-100 for UK stocks.
\\\\
\includegraphics[height=90mm,width=100mm]{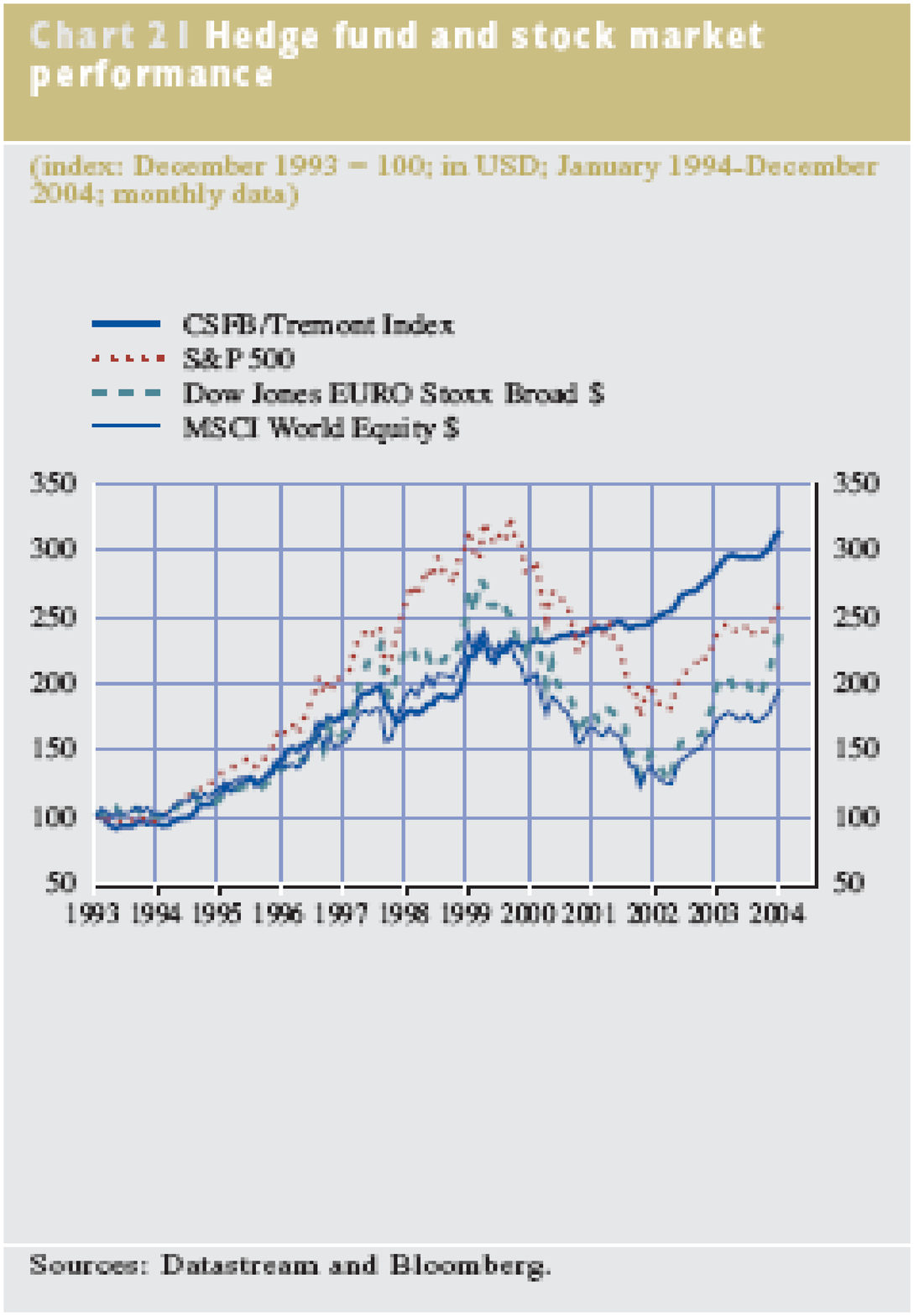}
\\\\
\subsection{Hedge Fund Performance Benchmark Targets}
With Mutual Funds only 1 type of performance benchmark typically
exists; the fund is expected to match or excel a particular index
e.g. FTSE-100 index, S\&P 500 index. This is an example of a
relative return target, which some Hedge Funds adopt as their
benchmark. However for Hedge Funds another benchmark exists called
absolute return targets.
\\\\
 An absolute return target is the typical
 benchmark choice for Hedge Funds and is the opposite of relative return. It is a
 fixed return target and the fund is expected to match/excel it regardless of
 the overall market performance.
Hedge fund managers use two main approaches to achieve absolute
return  targets: Market Timing and the Non-Directional approach.
\\\\
\textbf{Market Timing}\\ this approach takes positions by
anticipating the market trend or direction (either moving
up/down). This approach potentially offers high returns, as
demonstrated by Georg Soros in his Quantum Fund when speculating
on the British Pound in 1992.
\\\\
\textbf{Non-Directional}\\
An example of Non-Directional is A.Winslow's Hedge Fund; it is a
fund that eliminates some market risks, hence it can be considered
non-directional, whilst also benefitting from relative price
movements of assets. According to Fung \cite{FUNGPrimer} the
non-directional approach has evolved over the last decade and is
continuing to develop.

\subsection{Hedge Fund Organisation}
Hedge Funds typically prefer to concentrate their efforts on the
key activity of maximising investment return, so non-essential
operations are outsourced e.g. ``back office" functions. Actual
trading transactions too are outsourced to ``Prime Brokers". Prime
brokers are banks or securities firms, offering brokerage and
other financial services to large institutional clients e.g.
Pension Funds. It is also worth noting that Hedge Funds typically
reside ``offshore" to take advantage of more favourable tax
treatments and regulations.

\subsection{Fund of Hedge Funds (FOHF)}
A Fund of Hedge Funds is simplistically a Mutual Fund that invests
in multiple Hedge Funds e.g 15-25 different Hedge Funds,
furthermore F3 funds or fund of FOHF also exist. All these funds
provide diversification benefits and a method of investing in
Hedge Funds without requiring the skill to personally assess Hedge
Funds individually. Also, FOHF normally have significantly lower
minimum investment levels compared to a standard Hedge Fund, thus
increasing investment access to the general public.

\section{Hedge Fund Investment Strategies}
The investment strategies employed by various Mutual funds are
well documented, ranging from value investing to buying growth
stocks, with each having particular risk and return implications.
On the contrary, Hedge Fund investment strategies are far less
well documented and the variety of strategies are greater than for
Mutual Funds. Consequently, there is no widely accepted
categorisation of Hedge Fund strategies,for example, Stonham in
\cite{STONToo1} identifies 14 Hedge Fund strategy categories
whereas Fung \cite{FUNGPrimer} only has 7.
\\\\
We now describe the 7 main Hedge Fund investment strategies as
given by Fung \cite{FUNGPrimer}, which in turn are taken from MAR
(Managed Account Reports (one of the oldest sources of global
managed futures information )). The advantage of applying such
strategy categorisation is that different Hedge Fund return
characteristics can be explained by them (see \cite{FUNGPrimer}).

\subsection{Event Driven} An event driven strategy means a
position is taken to take advantage of price moves arising from
new market information release or events occurring. A good example
of such a strategy is to capitalise on merger and acquisition
announcements, which cause the target company's share price to
rise. An example is given below; Mark's and Spencer's share price
rose on announcement of a takeover by Philip Green at the end of
May 2004.
\\\\
\includegraphics[height=90mm,width=100mm]{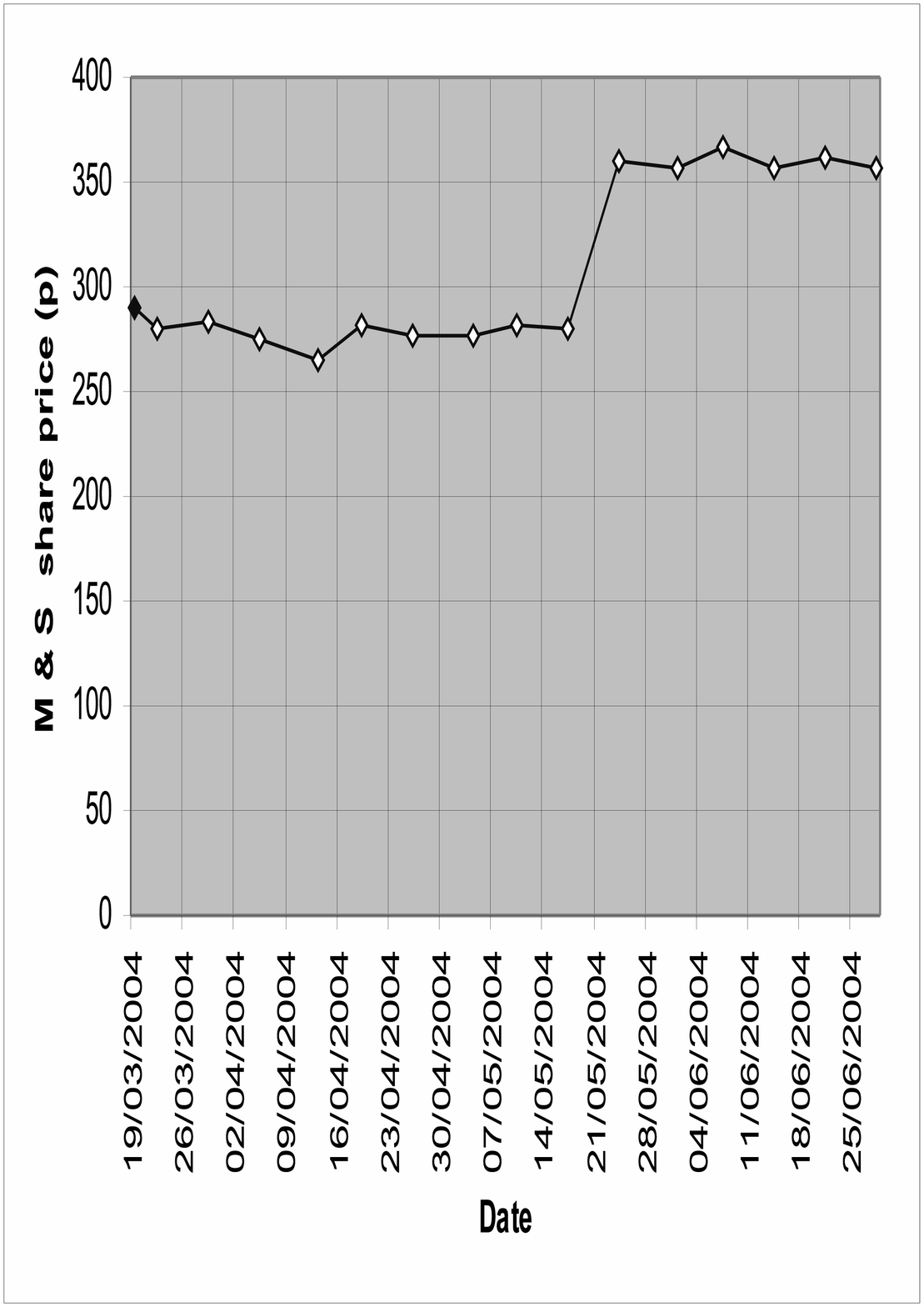}
\\
\subsection{Global} The Global strategy is an all-round category for funds
that invest in assets beyond those based in their home market.
Other than that, no more specific technique is associated with
this. A typical example would be a Hedge Fund investing in an
emerging market such as India.

\subsection{Global/Macro} The Global/Macro strategies utilise macroeconomic
analysis to capitalize on asset price changes that are strongly
linked to macroeconomics e.g. currencies, bonds, stock indices,
and commodities. As the name implies, this startegy is applied on
a global scale. For example, George Soros's Quantum Fund reputedly
made US\$1 billion in 1 day on September 1992 by speculating the
British Pound would exit the European Exchange Rate Mechanism.

\subsection{Market Neutral} Market neutral investment refers to funds
that hedge against market risk factors, thereby becoming "neutral"
to the market. This strategy profits by speculating on relative
price movements between assets or indexes. Examples of this method
include long-short equity, stock index arbitrage, fixed income
arbitrage. A good example of the long-short equity method is the
classic 1949 A.W. Jones Hedge Fund, who took long and short
positions in equities.

\subsection{Sector} Sector Hedge Fund investing concentrate on investing in specific
sectors e.g. airlines, telecoms, utilities sectors etc... . The
investment instrument itself can be a variety of types e.g. short
selling, long and leveraged positions.

\subsection{Short Selling and Long-Only}
Short selling and long-only Hedge Funds are those funds which will
\textsl{only} invest by shorting or going long respectively.

\newpage
\section{Hedge Fund Risk Models}
The necessity for Hedge Fund risk modelling and management
originates from 2 areas:
\begin{itemize}
\item Hedge Funds experiencing some of the greatest losses ever
witnessed by the investment community;
\item new regulatory
pressure enforcing more stringent Hedge Fund risk management.
\end{itemize}

Firstly, Hedge Funds have been responsible for numerous
catastrophic losses, causing them to completely collapse and
initiate a contagion effect by affecting numerous economic and
financial sectors. The most notorious example of such a
catastrophic loss being the Long Term Capital Management Hedge
Fund, which lost US\$2.1 Billion \cite{STONToo1} and almost
brought down the entire US financial system.
\\\\
Secondly, as already mentioned, Mutual Funds are tightly regulated
whereas Hedge Funds face little regulation. However, as Hedge
Funds have gained public attention and therefore more investment
interest, this along with spectacular Hedge Fund disasters have
prompted increased Hedge Fund regulation.
\\\\
It was not until after the 1997 Asian Currency Crisis though that
\textsl{regulators} became interested in regulating Hedge Fund
activities \cite{FUNGImpact}. The IMF (International Monetary
Fund) initiated a study on the market influence of Hedge Funds by
Eichengreen \cite{EICHHedge}. This study described Hedge Funds
activities and the potential problem of the market impact of Hedge
Funds.
\\\\
Moreover in 2004, the Securities and Exchange Commission now
required Hedge Fund managers and sponsors to register as
investment advisors under the Investment Advisor's Act of 1940.
This greatly increases the number of requirements placed on Hedge
Funds e.g. keeping records and creating a code of ethics. For more
information on SEC regulation visit the SEC website
$http://www.sec.gov/$.
\\\\
\includegraphics[height=230mm,width=150mm]{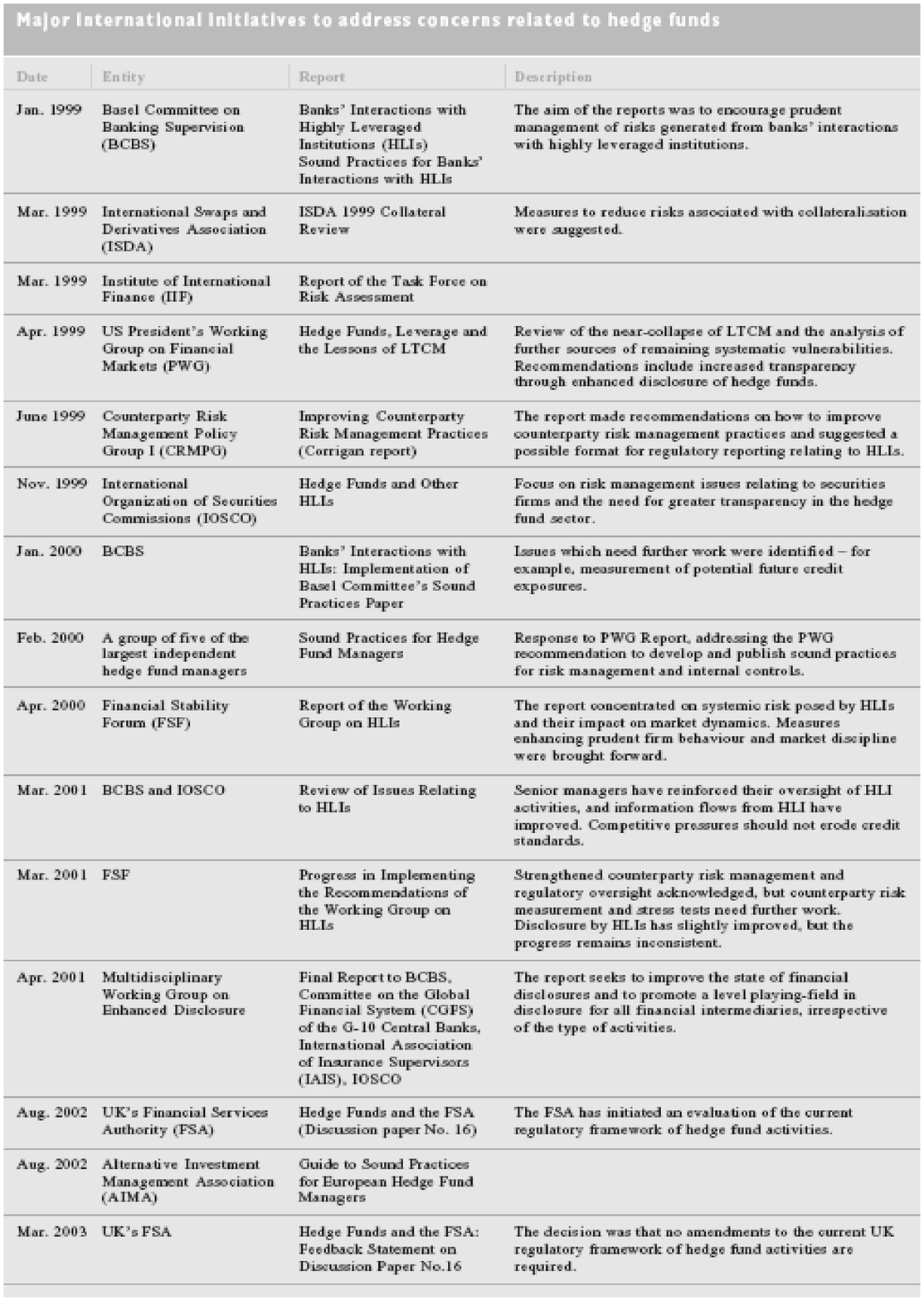}
\\\\
\includegraphics[height=100mm,width=150mm]{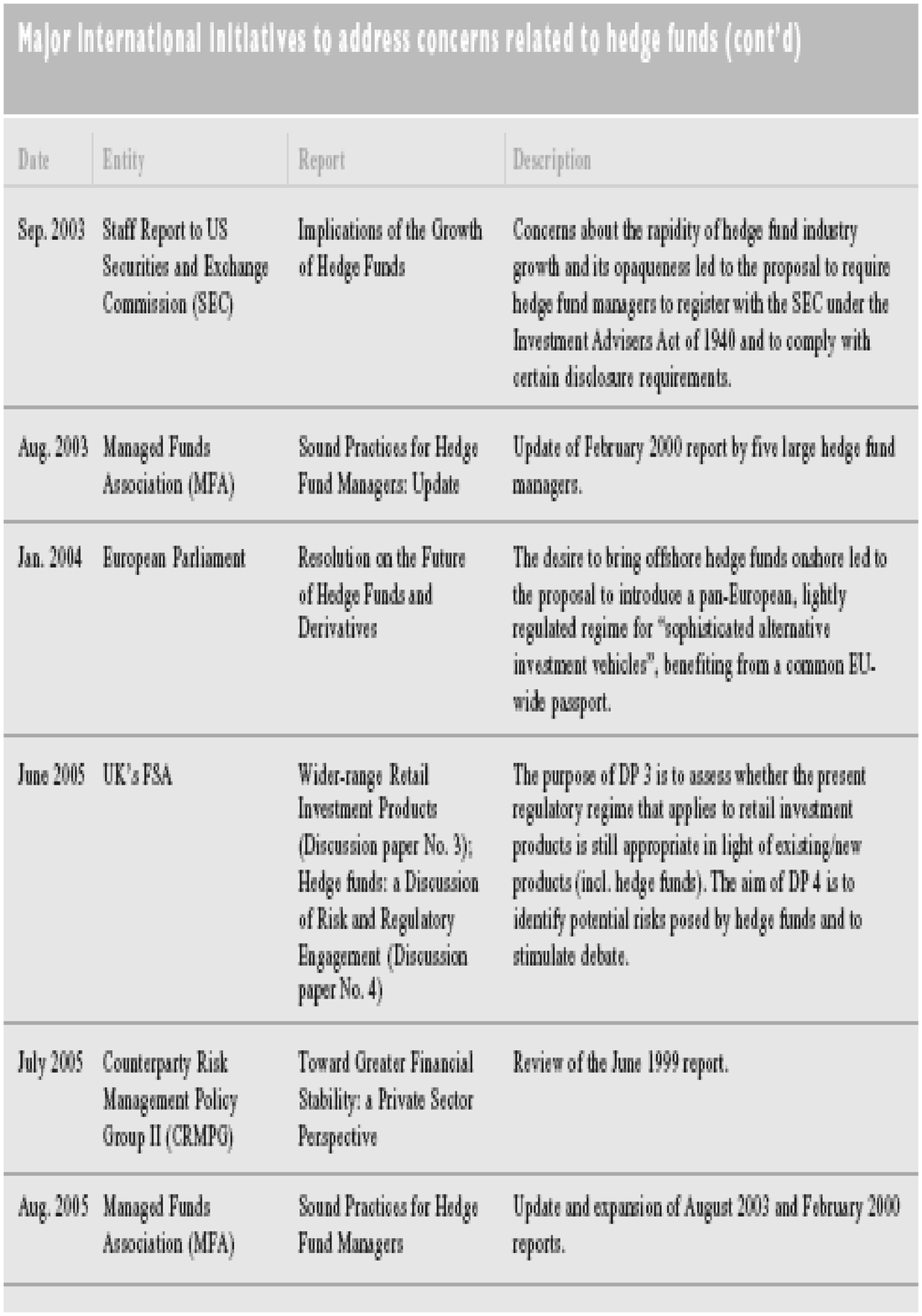}
\\\\
We now describe some of the quantitative risk models employed in
modelling Hedge Fund risks.

\subsection{Markowitz 's Portfolio Theory} Markowitz's
Portfolio Theory (from hereon MPT) is typically applied to
assets/portfolios whose return probability distributions
approximate to a Normal distribution. Although this approximation
is not strictly correct for Hedge Funds, it is still a workable
risk model. In fact Fung and Hsieh in \cite{FUNGMV} apply it to
rank Hedge Fund performances.
\\\\
Markowitz proposed a portfolio's risk is equal to the variance of
the portfolio's returns. If we define the weighted expected return
of a portfolio $R_{p}$ as
\begin{eqnarray}
R_{p}=\displaystyle\sum_{i=1}^{N}w_{i}\mu_{i},
\end{eqnarray}
then the portfolio's variance $\sigma_{p}^{2}$  is
\begin{eqnarray}\label{MPT portrisk}
\sigma_{p}^{2}=\displaystyle\sum_{i=1}^{N}\displaystyle\sum_{j=1}^{N}\sigma
_{ij}w_{i}w_{j},
\end{eqnarray}
where
\begin{itemize}
\item N is the number of assets in a portfolio;
\item i,j are the
asset indices and $i,j \in \{1,...,N \}$ ;
\item $w_{i}$ is the asset weight, subject to the constraints:
\\\\
$0 \leq w_{i}\leq 1$,
\\\\
$\displaystyle\sum_{i=1}^{N}w_{i}=1$;
\item $\sigma _{ij}$ is the covariance of asset i with
asset j;
\item $\mu_{i}$ is the expected return for asset i.
\end{itemize}
MPT also introduces the idea of an efficient frontier. For a given
set of funds or assets available to invest in, an upper concave
boundary exists on the maximum portfolio returns possible as risk
or variance increases. Furthermore this concave relation between
risk and return incorporates the theory of expected utility
concavely increasing with risk.
\\\\
\includegraphics[height=90mm,width=100mm]{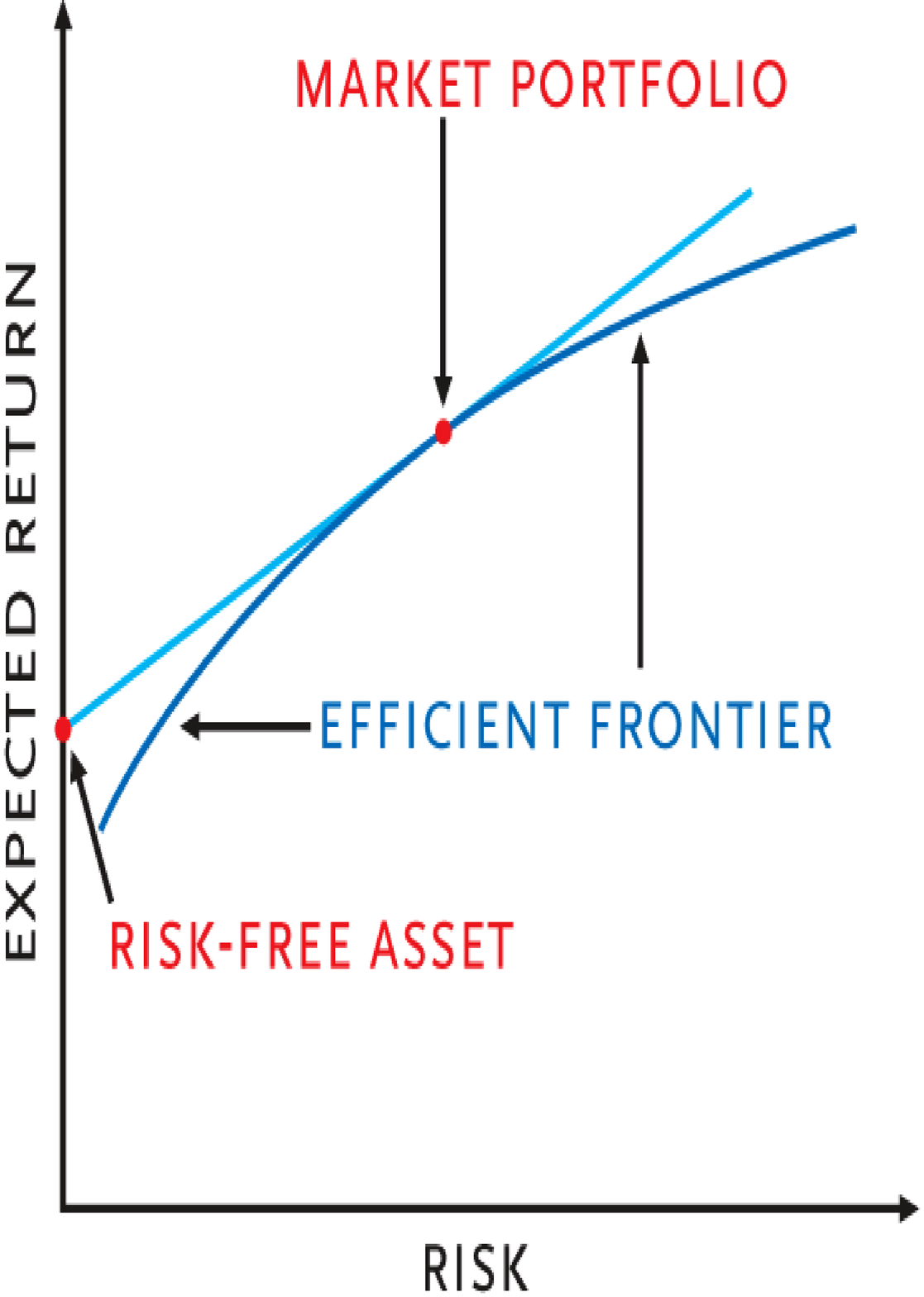}
\\\\
Notice that MPT shows that some funds can perform lower than the
risk free rate. Naturally one wishes to choose the market
portfolio which maximises return for a given level of
risk/volatility as shown.

\subsection{CAPM (Capital Asset Pricing Model)}
Capocci and Hubner \cite{CAPOAnal} state that in the 1980s CAPM
and its variants (e.g. Jensen's measure) were applied to Hedge
fund risk measurement. The CAPM model, based on MPT, was invented
by Sharpe \cite{SHARCapm}:
\[
    R_{a}=R_{f}+\beta (R_{m} -R_{f})+
    \varepsilon,
\]
where
\begin{itemize}
\item R$_{a}$ is expected return of an asset;
\item R$_{f}$ is the risk-free rate of return;
\item R$_{m}$ is the expected market return;
\item $\varepsilon$ is the error term;
\item $\beta=\dfrac{\sigma_{am}}{\sigma_{mm}}$;
\item $\sigma_{am}$ is the market and asset's covariance;
\item $\sigma_{mm}$ is the market's variance.
\end{itemize}
The CAPM model is applied generally in finance to determine a
theoretically appropriate return of an asset. It presumes that
investors must be compensated for investing in a risky asset in 2
ways 1)time value of money and 2)risk itself. The time value of
money is accounted for by the risk-free rate R$_{f}$ whereas the
return from risk arises from $\beta(R_{m} -R_{f})$. The term
$(R_{m} -R_{f})$ represents the expected risk premium, which is
the return obtained above the risk-free rate for investing in a
risky asset. The beta term can be considered the "sensitivity" of
the asset's risk to market risk (both measured by variance).
Consequently more "sensitive" assets ought to produce higher
returns by CAPM. The graph below shows how asset return is
linearly related to beta and that no beta implies a risk-free rate
of return.
\\\\
\includegraphics[height=50mm,width=100mm]{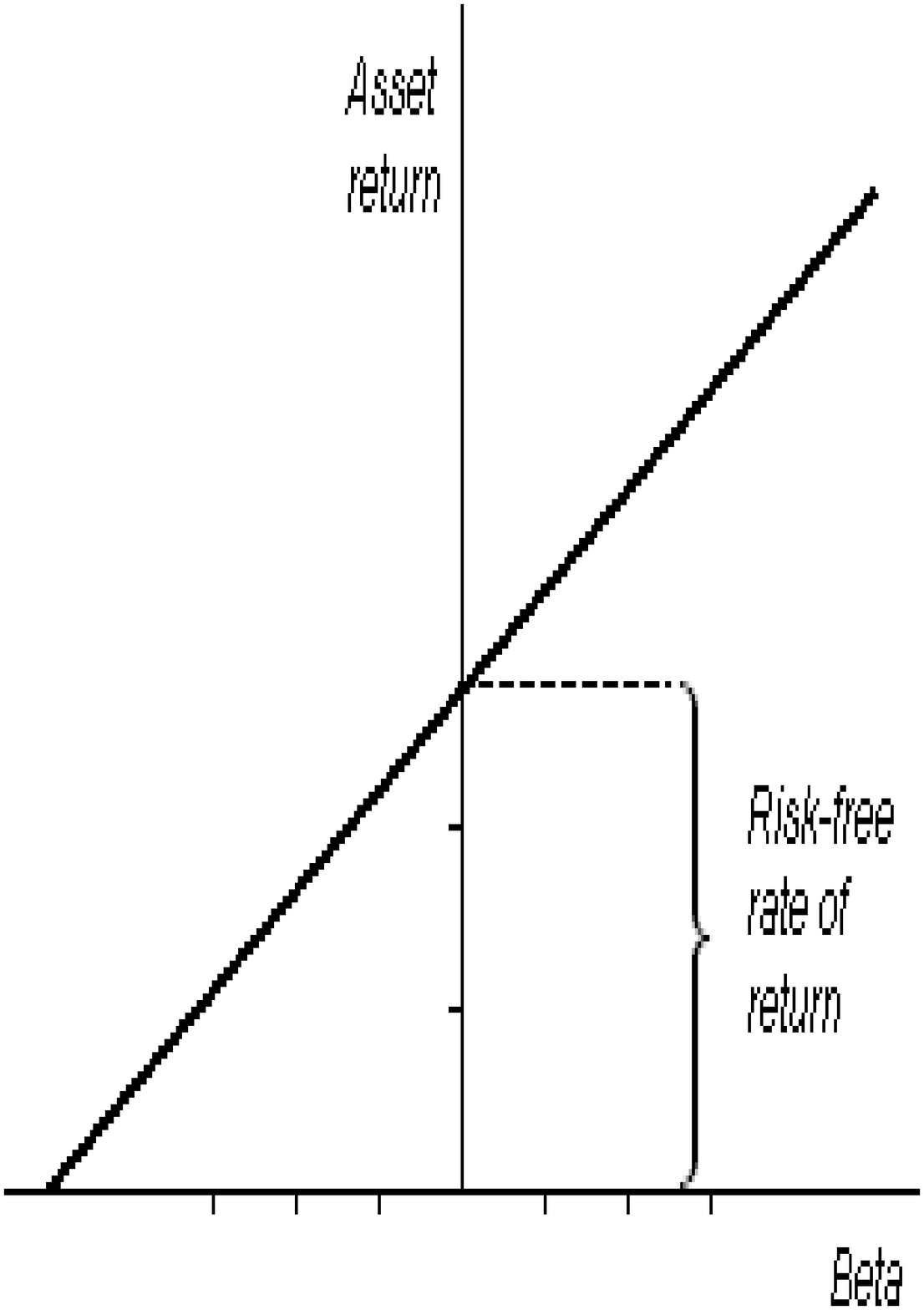}
\subsection{Sharpe Ratio and the Modified Sharpe Ratio}
The Sharpe Ratio $\mathcal{S}$, invented by Sharpe \cite{SHARPE},
is based on MPT's risk measure (variance):
\[
\mathcal{S}=\frac{R_{p}-R_{f}}{\sigma _{p}},
\]
where $\sigma _{p}$  is the portfolio return's standard deviation.
\\\\
The Sharpe ratio can be intepretted as "(Return - Risk-free
rate)/risk" since Sharpe considers standard deviation to be a risk
measure. The Sharpe ratio provides a portfolio risk measure in
terms of the quality of the portfolio's return at its given level
of risk.  A discussion on the Sharpe ratio can be found at
Sharpe's website (www.stanford.edu/~wfsharpe/).
\\\\
Fung and Hsieh in \cite{FUNGPerfH} and \cite{FUNGMV} use a
modified version of the Sharpe ratio to rank Hedge Fund
performance so to specifically cater for Hedge Fund return
distributions. This is simply the Sharpe ratio without subtracting
the risk free rate from the numerator:
\[
\text{Modified Sharpe Ratio=}\frac{R_{p}}{\sigma _{p}}.
\]

\subsection{Jenson's Alpha and Treynor ratio} Based on CAPM, Jensen
formulated a portfolio risk measure  to quantify portfolio returns
above that predicted by CAPM called $\alpha$:
\[
    \alpha=R_{p}-[R_{f}+\beta_{p}(R_{m} -R_{f})].
\]
One can interpret $\alpha$ as a measure of "excess returns"  or
portfolio manager's investment ability or i.e. "beating the
market".
\\\\
The Treynor ratio is a lesser well known portfolio ratio measure,
similar to the Sharpe ratio, but assesses portfolio performance on
a CAPM model basis:

\[
\text{Treynor Ratio=}\frac{R_{p}-R_{f}}{\beta_{p}}.
\]
Like the Sharpe ratio, the Treynor ratio can be interpretted as
the "quality" of portfolio return for the given level of risk but
risk measured on a CAPM theory basis.

\subsection{Three Factor Model of Fama and French} The CAPM model is a
single factor model that compares a portfolio with the market as a
whole. Fama and French modified this model in \cite{fama1993crf}
to take into account 2 empirical observations about asset classes
that tend to have higher returns:
\begin{itemize}
\item small sized companies;
\item value stocks (companies with high book to market value).
\end{itemize}
Having a higher return implies a higher risk premium associated
with them. The 3 factor model accounts for these higher premiums
with the following equation:
\[
    R_{a}=R_{f}+\beta_{p1}(R_{m} -R_{f})+\beta_{p2}SMB
    +\beta_{p3}HML+
    \varepsilon,
\]
where
\begin{itemize}
\item SMB is the difference in return for small and large sized
companies;
\item HML is the difference in return for high book to market value and low book to market value companies;
\item $\beta_{p1},\beta_{p2},\beta_{p3}$ are regression gradients (slopes).
\end{itemize}
Essentially the 3 factor model is a multiple linear regression
equation. Jagadeesh and Titman in \cite{JAGARetu} modify the CAPM
model by adding a momentum to account for return. Fung and Hsieh
in \cite{FUNGExtr} apply both these models to long/short equity
Hedge Funds, giving regression results.

\subsection{Sharpe's Asset Class Factor Model}
Sharpe in \cite{SHARAss} invented an asset factor model for risk
measurement of Mutual Funds but Fung and Hsieh in \cite{FUNGEmp}
have applied it to Hedge Funds. This model essentially suggests
that most Mutual Fund performances can be replicated by a small
number of major asset classes e.g. large capitalisation growth
stocks, large capitalisation value stocks, small capitalisation
stocks etc... . Using Fung and Hsieh \cite{FUNGEmp} notation
Sharpe's model is:
\[
R_{p}=\sum_{k}w_{k}F_{k} +\epsilon,
\]
subject to:
\begin{itemize}
\item $w_{k}=\displaystyle\sum_{j}x_{j}\lambda_{j}$;
\item $\epsilon=\displaystyle\sum_{j}x_{j}\epsilon_{j}$,
\end{itemize}
where
\begin{itemize}
\item j is the asset class;
\item k is the total number of asset classes;
\item $x_{j}$ is the weighting of asset class j;
\item $\lambda_{j}$ is the factor loading for asset j (change in fund
return/change in asset j return);
\item $\epsilon_{j}$ is the error term
for asset j
\end{itemize}
Thus Hedge Fund return is a weighted average of a small number of
asset classes, rather than a weighted average of a large number of
individual asset returns as in MPT.

\subsection{VaR (Value at Risk)}
VaR (value at risk) was invented by JP Morgan in 1994 as a general
risk management tool and has now become the industry standard for
risk. It has become a popular and important risk measure primarily
because of the Basel Committee, who standardise international
banking regulations and practises. Gupta and Liang in
\cite{GUPTVaR} applied VaR to Hedge Funds, specifically for
assessing a Hedge Fund's sufficient capital adequecy.
\\\\
VaR tells us in monetary terms how much one's portfolio can expect
to lose, for a given cumulative probability and for a given time
horizon. For example, for a cumulative probability of 99\% over a
period of 1 day, the VaR amount would tell us the amount by which
one would expect the portfolio to lose e.g.\$100.

VaR can be calculated by simulation using historical data or some
mathematical formula. VaR can also be calculated by the
``variance-covariance method" (also known as the delta-normal
method) but makes unrealistic assumptions about portfolio returns
e.g. returns are normally distributed.

\section{Problems with Hedge Fund Risk Modelling}
Most portfolio risk measures make unrealistic modelling
assumptions, particularly with respect to the assumed return
probability distributions for mutual funds. Risk measurement
assumptions become even more unrealistic for Hedge Funds. We now
explain the difficulties in Hedge Fund risk measurement.

\subsection{Non-Normal Return Distribution} As stated by Do et
al.\cite{DOHedge}, Hedge Fund returns do not approximate to normal
distributions, thus popular portfolio risk measures (which assume
a normal distribution) are inappropriate e.g. Sharpe ratio.
Furthermore Fung in \cite{FUNGPrimer} shows that the empirical
probability distribution of monthly returns for Hedge and Mutual
Funds differ significantly.
\\\\
Fung \cite{FUNGPrimer} proposes the reason for a non-normal return
distribution is a result of the diverse trading strategies
employed by Hedge Funds. Fung firstly suggests that Mutuals engage
in buy-and-hold strategies whereas Hedge Funds engage in much more
shorter term trading strategies. Secondly, Hedge Funds apply
substantial leveraging, whereas Mutuals have limited or strict
regulation on leveraging. Additionally, the relatively
regulation-free investment environment of Hedge Funds leads to
complex management strategies and high performance incentives
-these all affect Hedge Fund returns.

\subsection{Investment Strategy and Return Distribution} It has
been empirically observed that different investment strategies
significantly alter the return distribution, particularly the mean
and standard deviation. For example standard deviation, a common
risk metric, varies from a low 2.1\% in market neutral funds to
16.3\% in Global/Macro funds \cite{FUNGPrimer}. Consequently, it
has been argued it would be better to apply separate risk measures
for each Hedge Fund type (according to its strategy), rather than
treating all Hedge Funds as part of 1 homogenous class.

\subsection{Hedge Fund Failure Rate} Hedge fund survival rates
are significantly lower than other funds \cite{GARBECB} and
substantially vary; cumulative failure rates after 7 years range
from 32-66\% depending on the Hedge Fund's size. The table below
from the European Central Bank \cite{GARBECB} describes this:
\\\\
\includegraphics[height=90mm,width=80mm]{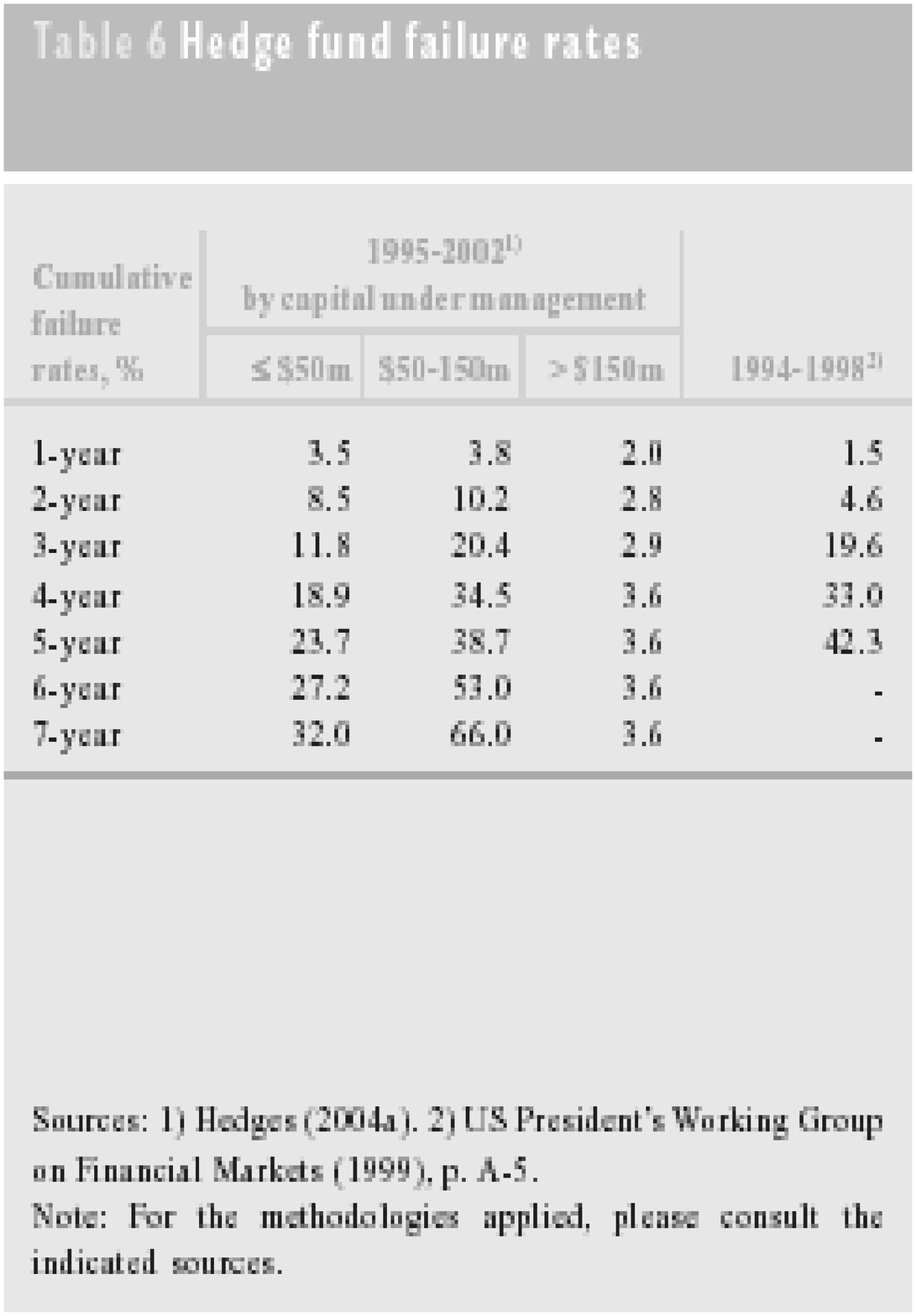}
\\\\
Thus the inclusion of non-existent Hedge Funds poses a problem
when assessing the overall performance of Hedge Funds (similar to
the survivorship bias issue with Mutual Fund performance).

\section{Hedge Funds Available On The Market}
\subsection{Close Man Hedge Fund}
Close Man Hedge Fund is an absolute return Hedge Fund. This fund
applies the market neutral investment strategy (specifically fixed
income arbitrage) by investing solely in Capital Guaranteed Bonds
issued by The Royal Bank of Scotland. Thus the fund is
theoretically insulated from market risks but can still benefit
from price movements using a variety of techniques. For this
particular fund, Close Man will engage in leveraging and using
swaps (a type of derivative) to boost returns.\\\\
See Close Man's website $http://www.closefm.com/$ for more detail.

\subsection{RAB Capital}
RAB Capital is a unique Hedge Fund in that it is one of the few UK
Hedge Funds (or more specifically FOHF) that is listed on the
London Stock Exchange (ticker symbol RAB.L).Their funds are
accessible to the general public rather than high net worth
individuals, although RAB warns ``These funds are not appropriate
for a novice investor". They specialise in a variety of absolute
return funds, some of which employ the long-only investment
strategy, where assets are bought on the basis that they are
considered undervalued.\\\\
See RAB Capital's website $http://www.rabcap.com/$ for more
detail.

\subsection{Thames River Capital}
Thames River Capital is an absolute return based Hedge Fund,
offering a range of regulated and unregulated funds. Each fund
uses various investment strategies, ranging from Global strategies
(see Global Emerging Market Fund) to market neutral strategies
using high leverage.\\\\
See Thames River Capital's wesbite
$http://www.thamesriver.co.uk/$ for more detail.

\subsection{Ikos Hedge Fund}
The founder and co-owner of her own hedge fund has made Elena
Ambrosiadou one of the richest women in Britain according to the
2006 Sunday Times Rich List. This hedge fund engages in ``program
trading" whereby trades are executed according to a computer
program. This method of trading has the advantage removing any
subjective decision making from speculation but can also result in
investments that one would strongly and intuitively disapprove.
Ikos focus on
exchange rate investing but also speculate in equities.\\\\
For more information on Ikos see $http://www.ikosam.com/$.

\section{Famous Hedge Funds Withdrawn From The Market}
All major funds are susceptible to collapsing, however, in the
case of Hedge Funds this is more frequent and the losses tend to
be substantially higher. It is therefore quite informative to
understand some of the spectacular Hedge Fund losses. We now
describe some Hedge Funds that were previously available on the
market but have now ceased trading.

\subsection{George Soros's Quantum Fund}
Perhaps the most famous Hedge Fund investor is Soros, who in 1 day
made US\$1 billion on September 6, 1992, by short selling the
British pound. In 1992, Britain was part of the ERM (European
Exchange Rate Mechanism) and Soros was able to anticipate the
currency devaluation of the British Pound. Consequently by
employing the Global/Macro investment strategy, Soros managed to
net a profit of US\$1 billion in 1 day. However years later, his
fund suffered massive losses; in 1998 Russia's defaulting crisis
created a loss of US \$2 billion.

\subsection{Long Term Capital Management (LTCM)}
Perhaps the most notorious Hedge Fund collapse was in September
1998; LTCM announced it had lost 44\% of its investors' capital in
August alone (US\$2.1 Billion) \cite{STONToo1}. For a detailed
case study of LTCM see Stonham in \cite{STONToo1},\cite{LTCM}.
\\\\
LTCM began trading with over \$1 billion of investor capital. LTCM
applied the Hedge Fund strategy of market neutral investment;LTCM
used the method of fixed income arbitrage, taking advantage of
temporary changes in prices. The market neutral strategy was
successful from 1994-98 but in 1998 Russian financial markets fell
into crisis. However, LTCM speculated that the situation would
quickly return back to normal again, so LTCM took large, unhedged
positions. Unfortunately, Russia began defaulting on its debts in
August 1998, causing LTCM to experience losses approaching \$4
billion as it was significantly exposed to Russian government
bonds. The US Federal Government then devised a rescue plan for
LTCM to avert a major US financial crisis and panic.

\subsection{Robertson's Tiger Management Fund}
Robertson's Hedge Fund invested by going long on undervalued
stocks whilst simultaneously short selling what he considered
overvalued stocks. For years this strategy was extremely
successful, giving annual returns of 43\% from 1980-86, so he
continued applying this strategy during the technology boom.
During the tech boom, Robertson rightly considered many stocks to
be overvalued and so began short selling such stocks with the
expectation overvalued stocks would eventually fall. Yet during
the tech boom a speculation bubble formed, causing the overvalued
stocks to continue to rise beyond expectation. Consequently
Robertson's fund collapsed in 2000 after heavy losses, just
\textit{before} the speculative bubble itself collapsed.

\section{The Case For and Against Hedge Funds}
Despite the potential to provide substantial returns, it would
appear conclusive that Hedge Funds ought to be abolished or at
least highly regulated. However the issue is far more complex than
one assumes. We now elaborate on the benefits and disadvantages of
Hedge Funds.

\subsection{The Case for Preserving Hedge Funds}
It can be argued Hedge Funds provide an economic benefit to
markets, in particular they aid price discovery. It has been
suggested that Hedge Funds take contrarian positions; they do not
engage in ``herd-mentality" trading, unlike Mutual Funds.
Therefore Hedge Funds buy or sell assets according to the
perceived fair value.
\\\\
A second economic benefit of Hedge Funds is that they aid
competition and the economic concept of the "invisible hand"
\cite{DANIHedge} and thrive on market inefficiencies. As traders
do not have instantaneous and costless access to market
information, asset mispricing or an arbitrage opportunities must
occur e.g. an asset trading in 2 different markets may have
different prices. Hedge Funds take advantage of such arbitrage
opportunities and so push prices to their no-arbitrage price.
\\\\
Another important economic benefit of Hedge Funds is liquidity
provision. Hedge Funds typically invest in riskier assets that
many investors would not consider. Hedge Funds therefore provide
much needed capital for investments.
\\\\
Hedge Funds can actually reduce overall risk rather than increase
it. Firstly, Hedge Funds take on riskier investments, thereby
``absorbing" some of the risk that would be concentrated in a
smaller number of funds. Additionally Hedge Funds are more willing
to invest in volatile markets, thereby ``absorbing" the effects of
market shocks.
\\\\
Hedge Funds are important as an investment product in itself. They
provide sophisticated investors with another vehicle for high
returns that would not be available in traditional Mutual Funds
\cite{DANIHedge}. They also provide diversification (a method of
reducing risk without reducing return by investing in more than 1
asset) as they represent a different investment class.
\\\\
A second benefit from a investor's perspective is that Hedge Funds
can provide "absolute" returns. Hedge Funds can achieve this
because they pursue a variety of sophisticated investment
strategies. Traditional Mutual Funds are limited in trading
strategies due to heavy regulation.

\subsection{The Case Against Hedge Funds}
Rather than aid market functioning, Hedge Funds have been
criticized for doing more harm than good. Firstly, rather than
contrarian investing, Hedge Funds engage in "herding"
\cite{DANIHedge}. Notable examples include the 1992 ERM crisis and
the 1997 Asian Currency Crisis.
\\\\
Secondly, it was suggested Hedge Funds provide much needed capital
by investing in risky assets, yet Hedge Funds have been blamed for
exhausting liquidity in the market \cite{DANIHedge}. Due to Hedge
Funds typically taking large positions and the trading strategies
they pursue, they are unable to make trades without causing a
massive price moves due to illiquidity (Fung supports this idea in
\cite{FUNGImpact}). Additionally, Hedge Funds are usually heavily
leveraged, increasing the likelihood of illiquidity e.g.LTCM.
However, Gupta in \cite{GUPTVaR} investigates capital adequacy
using VaR (value at risk) measures and concludes that
most Hedge Funds are adequately funded.\\
\\\\
Thirdly, Hedge Funds can prevent efficient market functioning by
causing market price distortions, rather than aiding price
discovery. Large volume trades can cause significant price
movements, rather than price movements occurring due to
company/economic fundamentals. Fung in \cite{FUNGImpact} cites
such examples as the 1992 ERM Crisis but concludes that Hedge
Funds overall do not distort prices beyond their company/economic
fundamentals.
\\\\
The Hedge Fund as a viable alternative investment product has also
been heavily disapproved. For instance some quotes from leading
academics on Hedge Funds:
\begin{itemize}
\item ``If you want to invest in something where they steal your
money and don't tell you what they`re doing, be my guest.", Eugene
Fama.

\item ``If there's a license to steal, it's in the hedge fund arena",
Burton Malkiel.

\end{itemize}

In an article in Forbes (May 14, 2004) Bernard Condon claims that
``You would do better giving your money to a monkey" than
investing in Hedge Funds. As a managed investment product Hedge
Funds command the highest management fees, typically around 20\%,
compared to mutual funds that normally charge around 1\%.
Additionally Hedge Fund investors have tougher withdrawal
constraints.
\\\\
Secondly as Fama mentions, Hedge Funds have poor transparency.
Regulatory bodies such as the SEC do not dictate the same strict
rules for Hedge Funds that it does for Mutual Funds: there are no
rules on publishing records on asset holdings and financial
performance, lack of transparency increases the chances of
investors being unable to effectively assess risk.
\\\\
Finally, Hedge Funds have a higher failure rate than Mutual Funds
and thus a higher credit risk. Hedge Fund face less regulation on
leveraging and investment strategies, thus are susceptible to a
higher probability of default e.g. LTCM. Consequently there is
less likelihood of capital recovery.

\section{Conclusion}
Hedge Funds are clearly a complex and unique investment product
that can produce extraordinary gains as well as losses. They have
and continue to thrive on the unregulated aspects of the business,
spawning a variety of innovative investment techniques. It has
only been in the past 10 years that regulatory bodies have
focussed on Hedge Fund regulation to avert previous Hedge Fund
disasters e.g. LTCM.
\\\\
Despite the clear necessity to understand such a powerful
investment, knowledge and understanding of the Hedge Fund industry
remains relatively poor. There is no consensus on the specific
definition of a Hedge Fund, very little literature is devoted to
Hedge Fund risk modelling and their various investment techniques.
Consequently there is a large scope for future research into Hedge
Fund risk management.

\newpage
\bibliographystyle{plain}
\addcontentsline{toc}{section}{References}
\bibliography{Ref}

\end{document}